# VizIt: A multi-view framework for exploring single-cell, spatial, and genetic data online

Chenhang Christopher Zhang[1,2,3,4], Yanqing Lou[5], Jie Yuan[1,2,3], Mingming Lu[1,2,3], Jacob Parker[1,2,3], Himanshu Chintalapudi[1,2,3], Zechuan Lin[1,2,3], Clemens R. Scherzer[1,2,3,6], Yuxuan Hu[1,2,3], Ruifeng Hu[1,2,3,#], and Xianjun Dong[1,2,3,7,#,*]

[1] Aligning Science Across Parkinson's (ASAP) Collaborative Research Network, Chevy Chase, MD.

[2] Adams Center for Parkinson's Disease Research, Yale School of Medicine, New Haven, CT.

[3] Department of Neurology, Yale School of Medicine, New Haven, CT.

[4] Concord Academy, Concord, MA.

[5] College of Engineering, Northeastern University, Boston, MA.

[6] Department of Genetics, Yale School of Medicine, New Haven, CT.

[7] Department of Neuroscience, Yale School of Medicine, New Haven, CT.

# equal contribution

*Correspondence should be addressed to: Xianjun Dong, PhD (xianjun.dong@yale.edu)

## Abstract

Multi-omic studies increasingly require data to be examined from complementary biological perspectives, yet interactive exploration remains fragmented across modalities and tools. We present VizIt, an open-source framework for multi-view exploration of single-cell and spatial transcriptomic, epigenomic and genetic data. VizIt connects gene-, cell type-, condition-, spatial-, genomic region- and variant-centered views, enabling seamless navigation across biological perspectives. We demonstrate VizIt through the Parkinson's Cell Atlas, a customizable interactive multi-omic resource.

## Introduction

Single-cell and spatial omics technologies increasingly enable biological systems to be characterized across complementary molecular dimensions, including gene expression, chromatin accessibility, spatial organization and genetic regulation[1–4]. As datasets become larger in size, higher resolution, and more multimodal, however, their interpretation presents a distinct challenge: researchers often need to examine the same data from multiple biological perspectives, in order to get a more comprehensive view (i.e. like the blind men touch the elephant analogy). A gene may be interrogated for its expression across cell types, a cell type for its marker genes or disease-associated variants, or a genetic variant for its regulatory effects across genes and cell types. These questions are biologically interconnected, yet their visualization often remains fragmented across separate plots, datasets and modality-specific tools.

Existing platforms address important aspects of this problem but were developed with different objectives. Analysis frameworks such as Seurat[5], Scanpy[6], and ArchR[7] provide extensive computational functionality but require programming expertise and computational environments. CZ CELLxGENE Discover and the Broad Single Cell Portal provide accessible exploration of centrally hosted single-cell datasets[8,9]. ShinyCell generates interactive application from individual single-cell dataset based on R shiny[10], where Vitessce offers flexible visualization components that can be incorporated into customized web applications[11]. gEAR provides a self-hostable environment for multi-omic data visualization and analysis[12]. Together, these resources have substantially improved access to single-cell data, but there remains a need for a readily deployable framework that can be organized diverse processed omics datasets around interconnected biological views, including genetic-association data alongside single-cell and spatial measurements.

Here we present VizIt, an open-source framework for multi-view exploration of single-cell, spatial, transcriptomic, epigenomic and genetic data. Rather than organizing exploration primarily around individual assays or predefined plots, VizIt allows the same datasets to be interrogated from multiple biological viewpoints and enables users to navigate between them. These include gene-, cell type-, condition-, genomic region-, peak- and variant-centered views. VizIt can be deployed as a standalone application or as a customizable institutional web portal, providing interactive access to processed omics datasets without requiring end users to write code.

## Multi-view exploration of omics data

The design of VizIt is motivated by a simple principle: **the same omics dataset can reveal different information when viewed from different biological perspectives**. In single-cell RNA-sequencing data, for example, a researcher may begin with a gene and ask in which cell types it is expressed. From the cell-type perspective, the corresponding question may instead be which genes characterize that population. If samples represent different diagnoses, treatments, ages or other covariates, a third perspective is whether expression within that cell type differs across conditions. These are not independent analyses but complementary views of the same underlying dataset.

VizIt implements these perspectives as interconnected views. For single-cell transcriptomic datasets, users can explore cell populations using interactive dimensionality-reduction representations such as UMAP or t-SNE, coloring cells by gene expression, sample characteristics, disease status or cell annotations. A gene-centered view displays its expression across cell populations and experimental conditions, whereas a cluster-centered view provides marker genes and differential expression results for the selected population. Navigation between views allows users to move from an observed cellular pattern to the genes underlying it, or from a gene of interest back to the cell populations and conditions in which it is altered.

The multi-view concept extends beyond transcriptomics. For single-cell ATAC-seq and multiome datasets, genomic signals can be interrogated from the perspective of a peak, gene, cell type or genomic interval. VizIt displays chromatin-accessibility profiles across cell types and genomic regions and allows these signals to be considered together with gene annotations and disease-associated variants. Users can pan and zoom across genomic intervals and configure displayed tracks, enabling regulatory features to be examined in their local genomic context.

Genetic regulatory associations provide a particularly useful example of why reciprocal views matter. VizIt supports expression and chromatin-accessibility quantitative trait locus results, including eQTLs and caQTLs. From a **gene-centered view**, users can examine variants associated with a selected gene across cell types. From a **variant-centered view**, they can instead ask which genes or regulatory features are associated with a selected variant. Individual variants, genes or peaks can then be followed into the corresponding reciprocal view. Genome-wide association study (GWAS) summary statistics can be displayed alongside these molecular associations, connecting disease-associated loci with cell-type-specific regulatory signals and candidate target genes.

Thus, rather than treating transcriptomic, epigenomic and genetic visualizations as independent endpoints, VizIt enables users to traverse related biological entities:

**cell type ↔ gene ↔ regulatory element ↔ genomic region ↔ genetic variant**

The precise relationships available depend on the processed data supplied to VizIt, but the common interface allows users to approach a biological question from different starting points and move between complementary views. Individual views can also be represented by URLs, facilitating the sharing of a specific biological context with collaborators.

## Extending multi-view exploration into spatial context

Spatial transcriptomics introduces another biological viewpoint: **where** molecular and cellular states occur within tissue. VizIt supports datasets generated by platforms including 10x Genomics Visium, Xenium and MERFISH. Gene expression and metadata can be visualized on tissue images or spatial coordinate maps, allowing molecular patterns to be related to anatomical compartments and local tissue organization.

The spatial view complements rather than replaces the cell- and gene-centered views. For example, a transcriptional signature identified in a single-cell dataset can be considered in terms of its spatial distribution, while genes of interest can be inspected across anatomical locations and cellular populations. Multiple datasets can also be opened for side-by-side inspection, allowing researchers to compare samples, donors, conditions, species or modalities. VizIt does not itself perform computational integration between these datasets; instead, it provides an interactive environment for examining processed results generated by upstream analytical workflows.

This distinction is important. VizIt is **not intended to replace analytical frameworks** such as Seurat, Scanpy, ArchR or specialized spatial-analysis software. Rather, it complements them by transforming processed outputs into navigable biological views that can be explored by computational and experimental researchers alike. This separation also enables analytical workflows to evolve independently of the visualization layer.

## From individual datasets to customizable multi-omic atlases

VizIt was designed not as a single centrally hosted repository but as a generalized framework that research groups can deploy for their own data collections. Processed datasets can be placed on the backend of a deployed server, and supported data can also be accessed from remote locations. The frontend retrieves data on demand and uses interactive browser-based rendering to support responsive exploration of embeddings, genomic tracks and spatial images.

A deployment can contain multiple datasets spanning different assays and studies while providing a common interface for navigating them. The organization and home interface can be customized for a particular research program or biological domain, allowing VizIt to serve as the foundation for project-specific or community-facing data portals.

VizIt can also be used locally as a standalone application, enabling investigators to interact with datasets without establishing a public portal. Conversely, deployment on institutional infrastructure provides multi-user web access and enables investigators to share datasets and individual views with collaborators. This architecture is intended to reduce the technical barrier between computational analysis and downstream biological interpretation.

## A common interface for changing biological perspectives

Visualization is often treated as the final stage of an omics workflow—a means of presenting results after an analysis has been completed. In practice, interactive visualization also participates in the analytical process. Observing expression of a disease-associated gene in a particular cell population leads naturally to questions about differential expression in that population; identifying a GWAS locus leads to questions about its regulatory targets; identifying a regulatory association leads to questions about the cell types in which it occurs; and a cell-type-specific molecular signal may prompt examination of its spatial distribution.

VizIt was developed to make these transitions straightforward. Its principal contribution is therefore not a new dimensionality-reduction algorithm or statistical method, but a **multi-view framework that connects processed single-cell, spatial, epigenomic and genetic data through the biological entities researchers use to formulate questions**.

By enabling investigators to move among gene-, cell type-, condition-, genomic-region-, regulatory-element- and variant-centered perspectives, VizIt provides a common interface between computationally generated results and biological interpretation. We anticipate that this multi-view paradigm will be particularly useful as single-cell studies increasingly combine molecular phenotypes, spatial information and human genetic associations within the same biological investigation.

## An implementation using ASAP Parkinson's disease datasets

As a real-world implementation, we used VizIt to build the **Parkinson's Cell Atlas**, a web resource hosting single-cell and spatial omics datasets from Parkinson's disease brain studies. The portal currently serves 13 datasets spanning single-nucleus RNA-seq, single-cell ATAC-seq, 10x Visium spatial transcriptomics, MERFISH, and single-cell eQTL and caQTL summary statistics. The atlas demonstrates how heterogeneous datasets can be organized within a domain-specific portal while retaining gene-, cell-,

spatial- and genomic-centered modes of exploration. The same framework can be adapted to other research areas without requiring the development of a new visualization application for each collection.

The datasets served by the Parkinson's Cell Atlas originate from the ASAP Parkinson Cell Atlas in 5D (PD5D) collection and from the ASAP Human Postmortem-Derived Brain Sequencing (PMDBS) collection (see **Table 1** below). Using these source datasets, we generated a unified set of interactive outputs in VizIt spanning several levels of analysis. For single-cell and spatial datasets, the portal provides sample-level UMAPs together with cell-cluster and gene-expression views; QTL datasets are presented through gene- and variant-centered association views; snRNA-seq and scATAC-seq datasets can be explored through genomic region views. These outputs allow the same underlying PD5D and PMDBS datasets to be examined across samples, genes, cell populations, variants, and genomic loci within a common interface. Representative outputs generated from the deposited datasets are shown in **Figure 1**.

**Table 1. Datasets used in Parkinson's Cell Atlas.**

| DATASET | ASSAY | BRAIN REGION | DOI LINK |
|---|---|---|---|
| PD5D_MTG_snRNAseq | snRNA-seq | Middle temporal gyrus | 10.5281/zenodo.16751625 |
| PD5D_MTG_VisiumST | 10x Visium ST | Middle temporal gyrus | 10.5281/zenodo.17242087 |
| PD5D_MTG_scATACseq | scATAC-seq | Middle temporal gyrus | pending |
| PD5D_Midbrain_snRNAseq | snRNA-seq | Midbrain | 10.5281/zenodo.18989317 |
| PD5D_Midbrain_scATACseq | scATAC-seq | Midbrain | pending |
| PD5D_MTG_sc_eQTL | Single-cell eQTL | Middle temporal gyrus | 10.5281/zenodo.13356892 |
| PD5D_Midbrain_sc_eQTL | Single-cell eQTL | Midbrain | 10.5281/zenodo.21937672 |
| PD5D_Midbrain_sc_caQTL | Single-cell caQTL | Midbrain | 10.5281/zenodo.21937672 |
| PMDBS_snRNAseq | snRNA-seq | MFG, hippocampus, substantia nigra | 10.5281/zenodo.16744323 |
| PMDBS_MFG_snRNAseq | snRNA-seq | Middle frontal gyrus | 10.5281/zenodo.16744323 |
| PMDBS_Hip_snRNAseq | snRNA-seq | Hippocampus | 10.5281/zenodo.16744323 |
| PMDBS_SN_snRNAseq | snRNA-seq | Substantia nigra | 10.5281/zenodo.16744323 |
| MERFISH_Demo | MERFISH ST | N/A (demonstration dataset) | N/A |

## Availability

VizIt is open source. Source code is archived at Zenodo (https://doi.org/10.5281/zenodo.21925072), and documentation is available at https://thedonglab.github.io/VizIt/. The Parkinson's Cell Atlas implementation is publicly accessible at https://pd5d.yale.edu/.

## Acknowledgements

The authors thank members of the Dong Lab for discussions and feedback related to this work. We also thank T. Liu, M. Feany, and J. Levin for input and feedback. This work was funded by Aligning Science Across Parkinson's [ASAP-000301] through the Michael J. Fox Foundation for Parkinson's Research (MJFF). For the purpose of open access, the author has applied a CC BY public copyright license to all Author Accepted Manuscripts arising from this submission.

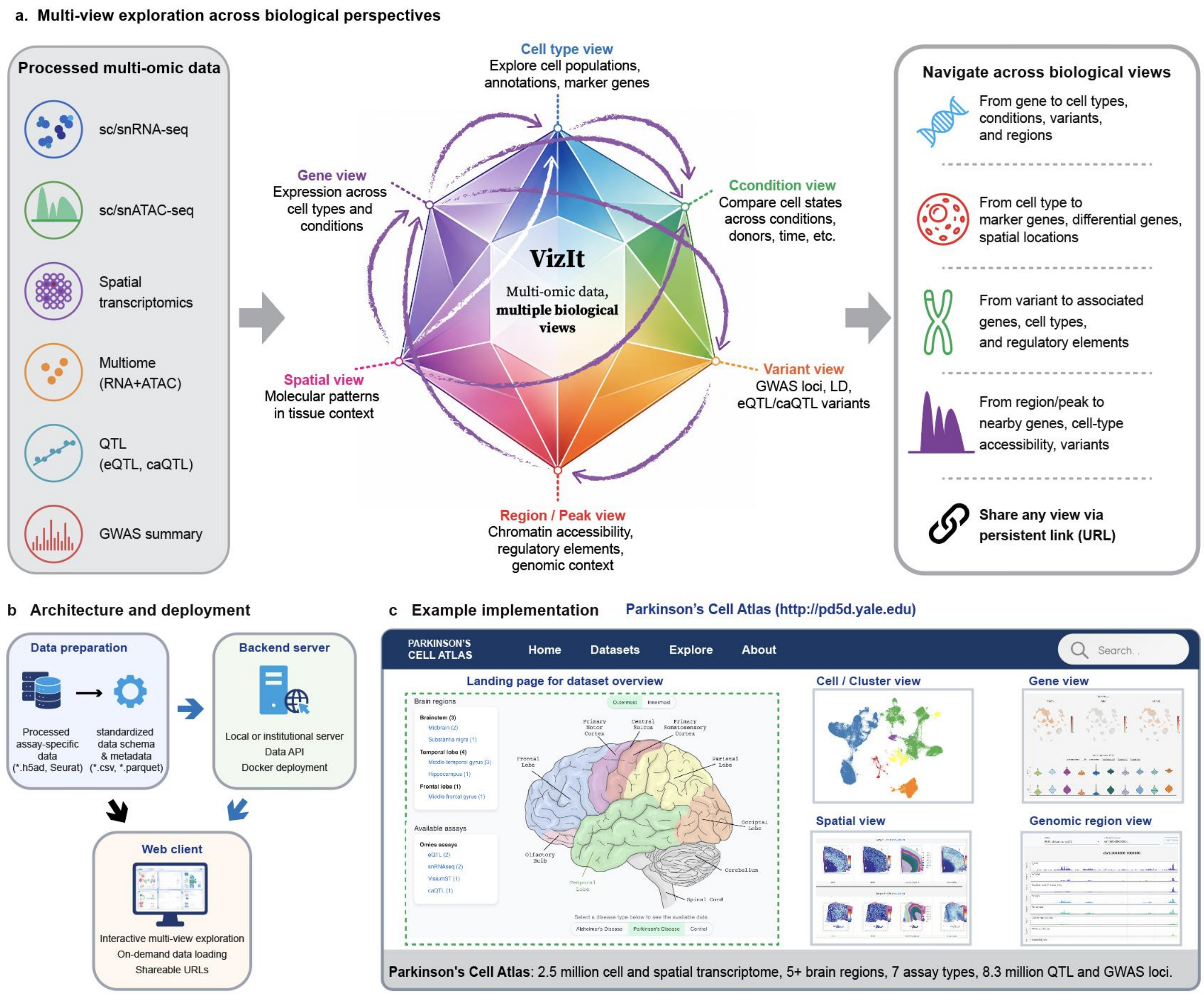

**Figure 1 | Multi-view exploration and implementation of VizIt.**

**a,** Multi-view exploration across biological perspectives. VizIt organizes processed multi-omic data, including single-cell or single-nucleus RNA-seq (sc/snRNA-seq), single-cell or single-nucleus ATAC-seq (sc/snATAC-seq), spatial transcriptomics, RNA-ATAC multiome, quantitative trait locus (QTL) and genome-wide association study (GWAS) data, into interconnected biological views. The same data can be explored from complementary gene-, cell type-, condition-, spatial-, variant- and genomic region/peak-centered perspectives. Users can navigate across these views—for example, from a gene to its expression across cell types and conditions, from a cell type to marker or differentially expressed genes and spatial locations, or from a genetic variant to associated genes and regulatory elements. Individual views can be shared through persistent URLs.

**b,** Architecture and deployment. Processed assay-specific data and metadata are standardized for deployment through a self-hosted VizIt backend server. The browser-based web client provides interactive multi-view exploration with on-demand data loading and shareable URLs. The backend can be deployed on local or institutional infrastructure using Docker.
**c,** Example implementation of VizIt as the Parkinson's Cell Atlas. The customized atlas provides access to millions of cells and spatial locations across multiple human brain regions and more than five assay types, together with cell-type-specific QTL and GWAS results. Representative interfaces show the dataset landing page and cell/cluster-, gene-, spatial- and genomic-region-centered views. It can be accessed at https://pd5d.yale.edu/.